\documentclass[twocolumn,preprintnumbers,amssymb,amsmath,aps,floatfix,prc,nofootinbib,superscriptaddress]{revtex4}
\usepackage{epsfig}
\usepackage{bm}
\usepackage{amssymb}
\usepackage{amsmath}
\usepackage{color}
\usepackage{subfigure}
\usepackage{hyperref}
\graphicspath{{./Figures/}}

\begin{document}
\title{Anisotropic flow, flow fluctuation and flow decorrelation in relativistic heavy-ion collisions: the roles of sub-nucleon structure and shear viscosity}

\author{Jie Zhu}
\email{zhujie@mails.ccnu.edu.cn}
\affiliation{Institute of Particle Physics and Key Laboratory of Quark and Lepton Physics (MOE), Central China Normal University, Wuhan, 430079, China}

\author{Xiang-Yu Wu}
\email{xiangyuwu@mails.ccnu.edu.cn}
\affiliation{Institute of Particle Physics and Key Laboratory of Quark and Lepton Physics (MOE), Central China Normal University, Wuhan, 430079, China}
\affiliation{Department of Physics, McGill University, Montreal, Quebec, H3A 2T8, Canada}

\author{Guang-You Qin}
\email{guangyou.qin@ccnu.edu.cn}
\affiliation{Institute of Particle Physics and Key Laboratory of Quark and Lepton Physics (MOE), Central China Normal University, Wuhan, 430079, China}

\begin{abstract}

We study the transverse momentum ($p_T$) differential anisotropic flow and flow fluctuation in Pb+Pb collisions at $\sqrt{s_{NN}}$=5.02 TeV at the LHC. 
A (3+1)-dimensional CLVisc hydrodynamics framework with fluctuating TRENTO (or AMPT) initial conditions is utilized to simulate the space-time evolution of the quark-gluon plasma (QGP) medium. 
The effects of shear viscosity and the sub-nucleon structure on anisotropic flow and flow fluctuation are analyzed. 
Our result shows that shear viscosity tends to suppress both flow coefficients (${v_2\{2}\}$, ${v_2\{4\}}$, ${\langle v_2\rangle}$) and flow fluctuation (${\sigma_{v_2}}$) due to its smearing effect on local density fluctuation. 
The flow coefficients appear to be insensitive to the sub-nucleon structure, whereas for flow fluctuation ${\sigma_{v_2}}$, it tends to be suppressed by the sub-nucleon structure in central collisions but enhanced in peripheral collisions. 
After taking into account the sub-nucleon structure effect, our numerical result can quantitatively describe the relative flow fluctuations (${v_2\{4\}/v_2\{2\}}$, $F({v_2})$) measured by the ALICE Collaboration at the LHC. 
We further investigate the effects of shear viscosity, sub-nucleon structure and initial condition model on the flow angle and flow magnitude decorrelations (${A_2^f}$, ${M_2^f}$) using the four-particle correlation method. 
We find that the flow decorrelation effect is typically stronger in central collisions than in peripheral collisions. 
The flow angle decorrelation is found to be insensitive to the shear viscosity and sub-nucleon structure, whereas the flow magnitude decorrelation shows quite different behavior when using TRENTO or AMPT initial condition model. 
Our study sheds light on the anisotropic flow, transport properties and initial structure of the QGP created in high-energy nuclear collisions.

\end{abstract}

\maketitle

\section{Introduction}\label{sec1}

One of the most significant discoveries in relativistic heavy-ion collisions performed at the Relativistic Heavy Ion Collider (RHIC) and the Large Hadron Collider (LHC) is the creation of the strongly-coupled Quark-Gluon Plasma (QGP). 
This novel hot and dense nuclear matter exhibits perfect fluid-like transport properties, as characterized by an extremely low ratio of shear viscosity to entropy density $({\eta_v/s})$~\cite{Adams:2003zg, Aamodt:2010pa, ATLAS:2011ah, Chatrchyan:2012ta}. 
Relativistic hydrodynamic approach~\cite{Wu:2021fjf,Pang:2018zzo,Schenke:2010rr,Schenke:2010nt,Schafer:2021csj,Karpenko:2013wva,Shen:2014vra} at high temperatures, combined with microscopic hadronic cascades~\cite{SMASH:2016zqf,Bass:1998ca,Du:2019obx} at low temperatures, have been very successful to describe the dynamical space-time evolution of heavy-ion collisions.

The strong (anisotropic) collective flow is one of the most important evidence for the formation of QGP, and has attracted significant attention over the past decades. 
It refers to the azimuthal anisotropy of the final produced particles, stemming from the initial state spatial anisotropy which converts to the final state momentum anisotropy as a result of the strong interaction among the QGP constituents~\cite{ Aguiar:2001ac, Broniowski:2007ft, Andrade:2008xh, Hirano:2009ah, Alver:2010gr, Petersen:2010cw, Qin:2010pf, Staig:2010pn, Teaney:2010vd}. 
The final-state anisotropies are typically observed by performing the Fourier analysis of the azimuthal angle distribution of the final state particles relative to the event-plane angle $\Psi_{{n}}$, i.e., $\frac{dN}{p_Tdp_T dy d\phi} \propto 1+2 \sum_{{n}=1}^{\infty} v_{{n}}(p_T,y) \cos \left[{n}\left(\varphi-\Psi_{{n}}(p_T,y)\right)\right]$, where the Fourier coefficient $v_n$ is called the $n$-th order harmonic flow. 
Early research has mainly focused on the elliptic flow $v_2$ because of the underlying elliptic geometry of the collision zone. 
Comprehensive studies~\cite{Song:2007ux, Song:2007fn,Schenke:2010rr,Schenke:2010nt,Ryu:2015vwa,Ryu:2017qzn,JETSCAPE:2020mzn,JETSCAPE:2020shq} have demonstrated that the magnitude of the elliptic flow $v_2$ is sensitive to the initial condition and transport properties of the QGP, such as specific shear viscosity ${(\eta_v / s)}$ and bulk viscosity ${(\zeta / s)}$. 
Recently, a Bayesian statistical method~\cite{JETSCAPE:2020mzn,JETSCAPE:2020shq, Parkkila:2021yha,Bernhard:2016tnd,Heffernan:2023utr} has been utilized to simultaneously constrain the initial conditions and transport coefficients by calibrating hydrodynamic calculations, e.g., the particle yield and flow magnitude, with the experimental data from the LHC and RHIC. 

In recent years, much attention has been paid to the initial state fluctuations in relativistic heavy-ion collisions, such as the fluctuations of nucleon positions or sub-nucleon structures inside nuclei, which can not only induce the odd-order harmonics flow  (e.g., $v_3, v_5$)~\cite{Alver:2010gr,Alver:2010dn,Qin:2010pf,Teaney:2010vd}, but also cause the fluctuations and correlations among flow magnitudes and flow angles from event to event in both transverse plane and longitudinal direction. 
Various flow observables have been defined to quantify flow fluctuations and correlations, such as event-plane correlations~\cite{Aad:2014fla, Bhalerao:2013ina, Qin:2011uw, Qiu:2012uy,Magdy:2022jai}, symmetric cumulants~\cite{ALICE:2016kpq, Zhu:2016puf, Giacalone:2016afq,Li:2021nas}, non-linear response coefficients~\cite{Yan:2015jma, Qian:2016fpi, Acharya:2017zfg, Giacalone:2018wpp} and $v_{n}$-$p_T$ correlation~\cite{PhysRevC.93.044908}. 
Compared to the flow coefficients, the flow fluctuation and correlation observables can provide additional constraints on the evolution dynamics and transport properties of the QGP. 
For example, symmetric cumulants are sensitive to the temperature dependence of shear viscosity $\eta_v/s(T)$~\cite{Parkkila:2021yha,ALICE:2016kpq,ALICE:2017kwu}, and $v_{n}$-$p_T$ correlation is sensitive to the initial nuclear structure~\cite{Giacalone:2021clp,Bally:2021qys,Giacalone:2020awm,Nijs:2020ors,Schenke:2020uqq,Fortier:2023xxy,PhysRevC.93.044908}.  

Another important set of flow observables is the factorization breaking (or the decorrelation effect) of the anisotropic flow~\cite{Pang:2018zzo,Pang:2015zrq,Pang:2014pxa,Bozek:2015bna,ATLAS:2020sgl,Bozek:2017qir,Wu:2018cpc,Sakai:2020pjw,Zhao:2017yhj,CMS:2015xmx}. 
A common method for studying the factorization breaking effect is to construct two-particle correlations using flow vectors.
Such method takes into account the combined contributions from flow magnitude and flow angle~\cite{ATLAS:2017rij,Bozek:2017qir,Bozek:2021mov,Wu:2018cpc,Jia:2017kdq}, 
The factorization breaking effect has been confirmed in Au+Au collisions at RHIC, as well as in Xe+Xe and Pb+Pb collisions at the LHC~\cite{ATLAS:2020sgl,CMS:2017xnj,ATLAS:2017rij,CMS:2015xmx,ALICE:2022dtx}. 
Recently, the ALICE Collaboration has developed new observables based on four-particle correlations, which can separately measure the contributions from flow magnitude and flow angle  in the transverse plane~\cite{ALICE:2022dtx}. 
Similar observables have also been defined by the ATLAS Collaboration to separate the contributions from flow magnitude and flow angle along the longitudinal direction~\cite{ATLAS:2017rij}. 
Distinguishing the contributions from flow magnitude and flow angle offers a unique opportunity to understand and constrain the fundamental properties of the QGP medium. 

Our paper is organized as follows. In Sec.~\ref{sec2}, we introduce the setup of the event-by-event (3+1)-dimensional CLVisc viscous hydrodynamics framework. In Sec.~\ref{sec3}, the $p_T$ differential flow observables are constructed based on two-particle and four-particle correlations. In Sec.~\ref{sec4}, we present our numerical results for the elliptic flow fluctuation for identified particles, and the angle and magnitude decorrelations of elliptic flow for charged particles. The summary will be presented in Sec.~\ref{sec4}.

\section{Hydrodynamics model of relativistic heavy-ion collision}\label{sec2}

In this work, we utilize the event-by-event (3+1)-dimensional CLVisc viscous hydrodynamic framework~\cite{Pang:2012he, Pang:2018zzo, Wu:2021fjf}, which includes the fluctuating TRENTO initial condition~\cite{Moreland:2014oya,Soeder:2023vdn,Ke:2016jrd}, the second-order dissipative hydrodynamic evolution and Cooper-Frye hadronic sampler, to simulate the space-time evolution of the QGP medium and hadron production in Pb+Pb collisions at $\sqrt{s_{NN}}$=5.02 TeV at the LHC.  

The theoretical descriptions of the initial states of the QGP created in relativistic heavy-ion collisions from the first principle remain a significant challenge. 
In this study, we employ the parameterized initial condition model, TRENTO~\cite{Moreland:2014oya,Soeder:2023vdn,Ke:2016jrd}, to phenomenally simulate the initial space-time distribution of the QGP fireball. 
In the TRENTO initial condition, the positions of nucleons are sampled following the Woods-Saxon distributions. 
Compared with the traditional Glauber model~\cite{Miller:2007ri}, the collisional probability of nucleons is determined by the effective parton cross-sections, which are tuned to fit the inelastic nucleon cross-section. 
Recently, the sub-nucleon degree of freedom has been incorporated into the TRENTO model, which provides a possible approach to investigate the substructure of nucleons and extra sources of flow fluctuations~\cite{Giacalone:2021clp,JETSCAPE:2020shq,Nijs:2021clz,Schenke:2020unx,Shen:2016zpp,Loizides:2016djv,Noronha-Hostler:2015coa,Mantysaari:2016jaz,Mantysaari:2016ykx}. Once the positions of wounded nucleons are determined, one can obtain the thickness functions of the projectile and target nuclei as follows:
\begin{equation}
T_{A/B}=\sum_{i=1}^{N_{A/B}} w T_p\left(x,y;x_i, y_i\right).
\end{equation}
Here $T_p$ is the thickness function originating from each wounded nucleon with Gaussian form, and $w$ is a random weight taking from a Gamma distribution to account for the multiplicity fluctuations in p+p collisions. Naturally, one may define the reduced thickness function by considering the generalized average of the thickness functions of two nuclei, 
\begin{equation}
T_R\left(p ; T_A, T_B\right) \equiv\left(\frac{T_A^p+T_B^p}{2}\right)^{1 / p},
\end{equation}
where the parameter $p$ characterizes the different approximations for the entropy deposition.
In the new version of the TRENTO initial condition model, the sub-nucleon structure is included. 
The positions of constituent partons are assumed to follow a Gaussian distribution within the nucleons, and the same entropy deposition function is used as the nucleons. 
In this work, we choose an IP-Glasma-like entropy deposition, i.e., $p=0$~\cite{JETSCAPE:2020mzn,JETSCAPE:2020shq}. 
The number of constituent partons is set to 3. 
The Gaussian widths are set as 0.5~fm for nucleons and 0.2~fm for partons. 
By utilizing the scale factor $K$, the initial transverse entropy profile is assumed to be, 
\begin{equation}
\left.\frac{d S}{d y}\right|_{\tau=\tau_0}=K \cdot T_R\left(p ; T_A, T_B\right),
\end{equation}
where the factor $K$ can be tuned to fit the final particle yield ${dN/d{\eta}}$ in the most central collisions. 
Additionally, we assume the initial proper time $\tau_0$ as 0.6~fm. 
For the longitudinal direction, we use the following plateau-like envelope function,
\begin{equation}
H(\eta)=\exp \left[-\frac{\left(\eta-\eta_{\text {flat }}\right)^2}{2 \eta_{{\rm gw}}{ }^2} \theta\left(|\eta|-\eta_{\text {flat }}\right)\right].
\end{equation}
In this work, we take the parameters ${\eta_{\rm flat}}=1.7$ and  ${\eta_{\rm gw}}=2.0$. 

After the initial entropy density distribution has been constructed, the (3+1)-dimensional CLVisc viscous hydrodynamics model is used to simulate the space-time evolution of the QGP fireball. The following energy-momentum conservation equation is solved: 
\begin{equation}
    \partial_{\mu} T^{\mu\nu} = 0,
\end{equation}
where $T^{\mu\nu}$ is the energy-momentum tensor of the QGP system. 
Since the QGP medium is not fully in local thermal equilibrium, we also solve the evolution of shear tensor and bulk pressure using the second-order Israel-Stewart equation~\cite{Denicol:2018wdp},
\begin{equation}
\begin{gathered}
\tau_{\Pi} D \Pi+\Pi=-\zeta \theta-\delta_{\Pi \Pi} \Pi \theta+\lambda_{\Pi \pi} \pi^{\mu \nu} \sigma_{\mu \nu} \\
\tau_\pi \Delta_{\alpha \beta}^{\mu \nu} D \pi^{\alpha \beta}+\pi^{\mu \nu}=\eta_v \sigma^{\mu \nu}-\delta_{\pi \pi} \pi^{\mu \nu} \theta+\tau_{\pi \pi} \pi^{\lambda\langle\mu} \sigma_\lambda^{\nu\rangle}\\+\varphi_1 \pi_\alpha^{\langle\mu} \pi^{\nu\rangle \alpha}.
\end{gathered}
\end{equation}
Here $D$ is the covariant derivative, and $\theta=\partial_{\mu}u^{\mu}$ is the expansion rate. The traceless symmetric tensor is defined as $\pi^{\langle\mu \nu\rangle}=\Delta_{\alpha \beta}^{\mu \nu}\pi^{\alpha \beta} $ with second-order symmetric projection operator $\Delta_{\alpha \beta}^{\mu \nu}=\frac{1}{2}\left(\Delta_\alpha^\mu \Delta_\beta^\nu+\Delta_\alpha^\nu \Delta^\mu{ }_\beta\right)-\frac{1}{3} \Delta^{\mu \nu} \Delta_{\alpha \beta}$. $\tau_{\pi}$ and $\tau_\Pi$  are relaxation times of shear tensor and bulk pressure. $\delta_{\Pi \Pi},\lambda_{\Pi \pi},\delta_{\pi \pi},\tau_{\pi \pi},\varphi_1$ are second-order transport coefficients. The values of relaxation time and transport coefficients are summarized in Table \ref{tab:relaxation}.

\begin{table}[h]
\begin{tabular}{|c|c|c|c|}
\hline\hline$\tau_{\Pi}$ & $\delta_{\Pi \Pi}$ & $\lambda_{\Pi \pi}$ & \\
\hline$\frac{\zeta(T)}{15\left(\frac{1}{3}-c_s^2\right)^2(e+p)}$ & $\frac{2}{3} \tau_{\Pi}$ & $\frac{4}{5}\left(1-c_s^2\right) \tau_{\Pi}$ & \\
\hline\hline$\tau_\pi$ & $\delta_{\pi \pi}$ & $\tau_{\pi \pi}$ & $\varphi_1$ \\
\hline$\frac{{\eta_v}}{T}$ & $\frac{4}{4} \tau_\pi$ & $\frac{5}{7}$ & $\frac{9}{70} \frac{4}{e+p}$ \\
\hline
\end{tabular}
\caption{The transport coefficients for the second-order Israel-Stewart equation of motion.}
\label{tab:relaxation}
\end{table}

In this work, the specific shear viscosity ${\eta_v}/s$ is taken as 0.16, the temperature dependence of bulk viscosity  ${\zeta/s(T)}$ is taken from Ref.~\cite{Bernhard:2016tnd}, and the equation of state is taken from the HotQCD Collaboration~\cite{HotQCD:2014kol}. 
When the temperature of QGP fluid cell drops down to the freezeout temperature ($T_\mathrm{frz}$=150 MeV), the momentum of the thermal hadrons is obtained using the Cooper-Frye prescription,
\begin{equation}
\frac{d N_i}{d Y p_T d p_T d \phi}=\frac{g_i}{(2 \pi)^3} \int p^\mu d \Sigma_\mu f_{\mathrm{eq}}(1+\delta f),
\end{equation}
where the thermal equilibrium term $f_{\rm {eq}}$ and out-of-equilibrium correction $\delta f$ are given as,
\begin{equation}
\begin{aligned}
f_{\mathrm{eq}} & =\frac{1}{\exp \left[\left(p \cdot u-\mu_i\right) / T_{\mathrm{frz}}\right] \pm 1}, \\
\delta f & =\left(1 \mp f_{\mathrm{eq}}\right) \frac{p_\mu p_\nu \pi^{\mu \nu}}{2 T_{\mathrm{frz}}^2(\varepsilon+P)}.
\end{aligned}
\end{equation}
In this work, we neglect the correction from bulk pressure due to its theoretical uncertainty. 
After the thermal hadrons are produced, they undergo resonance decay procedures. 
The centrality classes are determined from the total entropy at midrapidity at initial time $\tau_0$ within the TRENTO initial condition. 
In order to improve the statistics, we run 2000 hydrodynamics events and oversample 3000 times for each hydrodynamics event at each centrality bin.

\begin{figure*}[t]
	\centering
	\includegraphics[width=\textwidth]{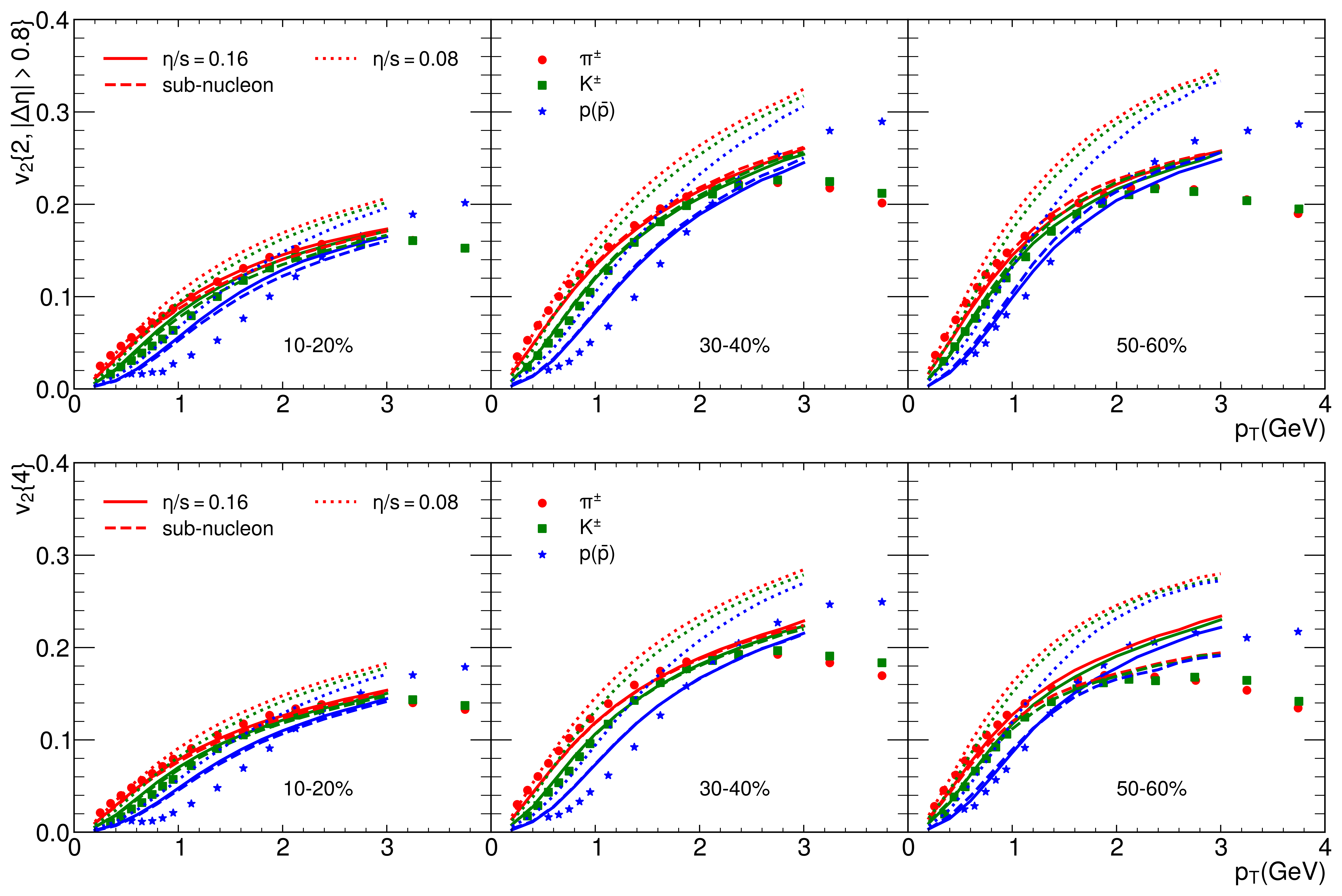}
	\caption{The elliptic flow, $v_2\{2\}$ (with $\vert \Delta {\eta} \vert > 0.8$) and ${v_{{2}}\{4\}}$, for identified particles  as a function of transverse momentum $p_T$  in Pb+Pb collisions at $\sqrt{s_{NN}}$ = 5.02 TeV for 10-20\%, 30-40\%, and 50-60\% centrality classes. 
  The solid  and dotted lines represent the results from hydrodynamics simulation with TRENTO initial condition for $\eta_v/s$=0.16 and $\eta_v/s$=0.08, respectively. The blue dashed line represents the hydrodynamics result with sub-nucleon structure in the TRENTO initial condition and $\eta_v/s=0.16$. The data are taken from ALICE collaboration~\cite{ALICE:2022zks}.} 
	\label{v24}
\end{figure*}

\begin{figure*}[t]
	\centering
	\includegraphics[width=\textwidth]{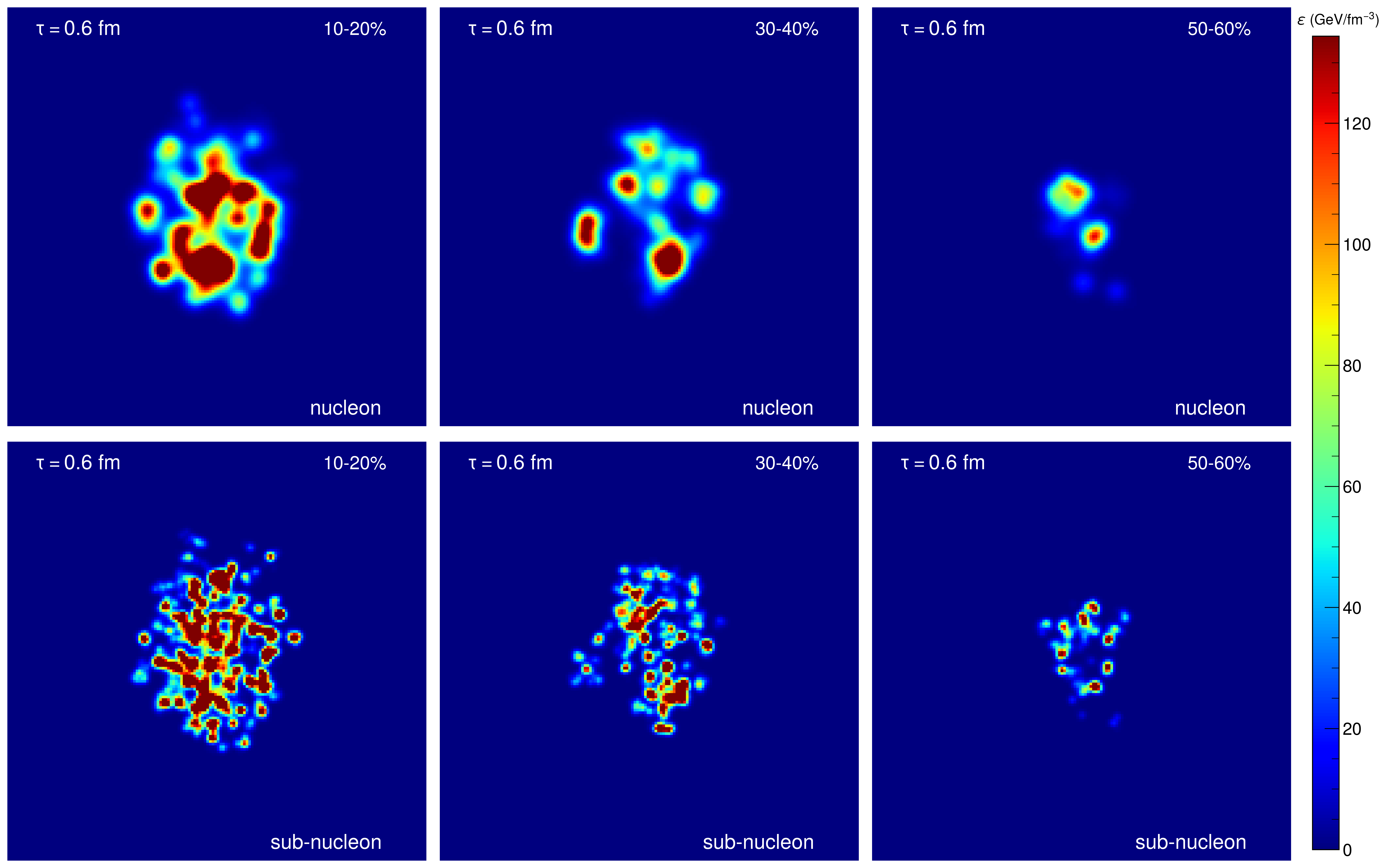}
	\caption{The initial energy density at $\tau_0$ = 0.6 fm and $\eta_s$ = 0 in Pb+Pb collisions at $\sqrt{s_{NN}}$=5.02 TeV. From left to right  are  0-5\%, 10-20\% and 30-40\% centrality classes. The upper panel stands for the initial conditions with the nucleon size 0.5 fm, while the bottom panel shows the initial conditions with the sub-nucleon structure (i.e., the nucleon consists of three constituent quarks with size 0.2 fm).}
	\label{ini-sub}
\end{figure*}

\begin{figure*}[t]
	\centering
	\includegraphics[width=\textwidth]{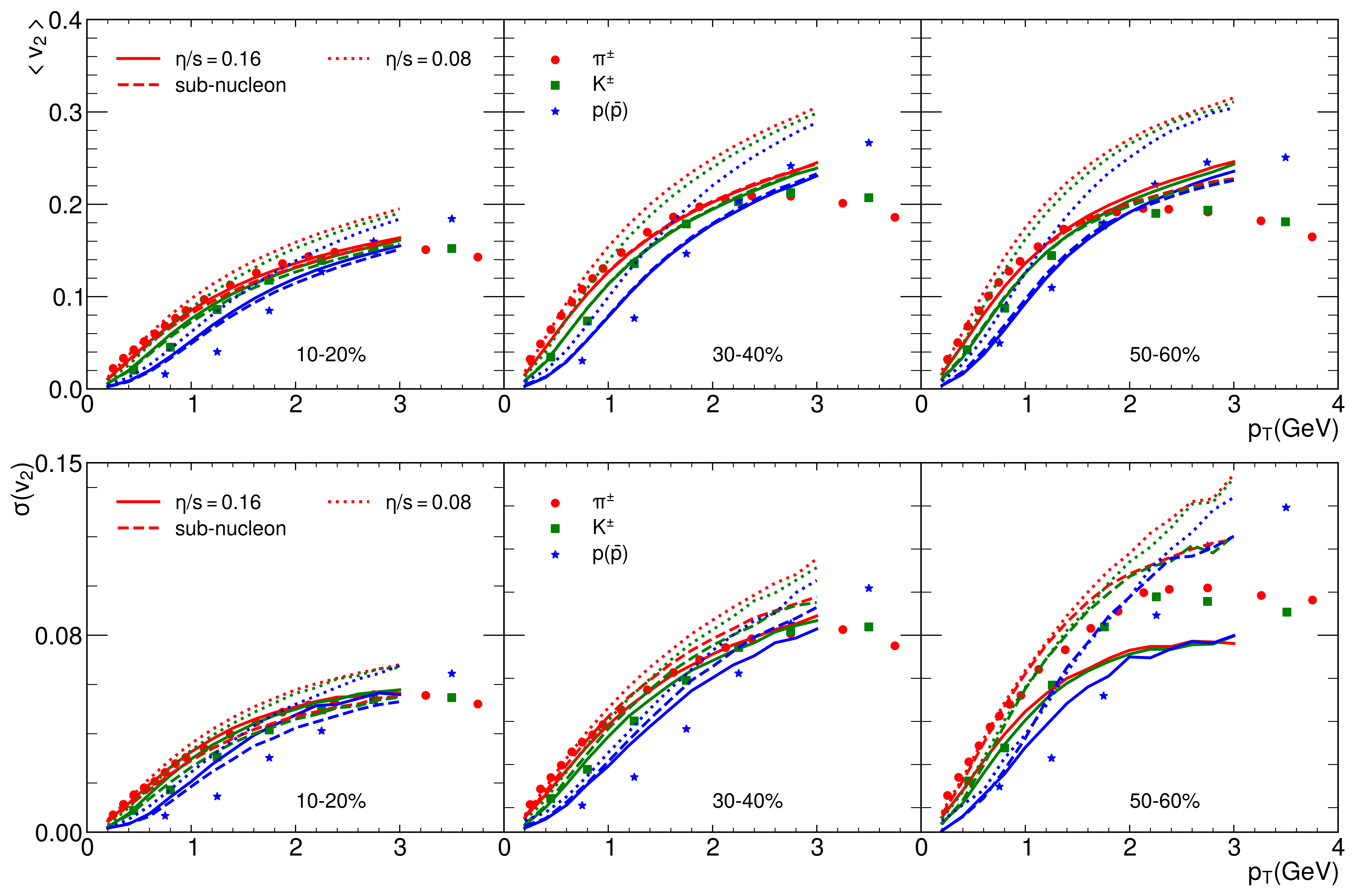}
	\caption{The mean flow $\langle {v_2} \rangle$ and flow fluctuation ${v_2(\sigma_{v_2})}$)  for identified particles as a function of  transverse momentum $p_T$  in Pb+Pb collisions $\sqrt{s_{NN}}$ = 5.02 TeV for 10-20\%, 30-40\%, and 50-60\% centrality classes. 
  The solid  and dotted line represent the hydrodynamic results with TRENTO initial condition for $\eta_v/s$=0.16 and $\eta_v/s$=0.08, respectively. The blue dashed line represents the hydrodynamics result with the sub-nucleon structure in the TRENTO initial condition and $\eta_v/s=0.16$. The data are taken from ALICE collaboration~\cite{ALICE:2022zks}.}
	\label{sigma}
\end{figure*}

\begin{figure*}[t]
	\centering
	\includegraphics[width=\textwidth]{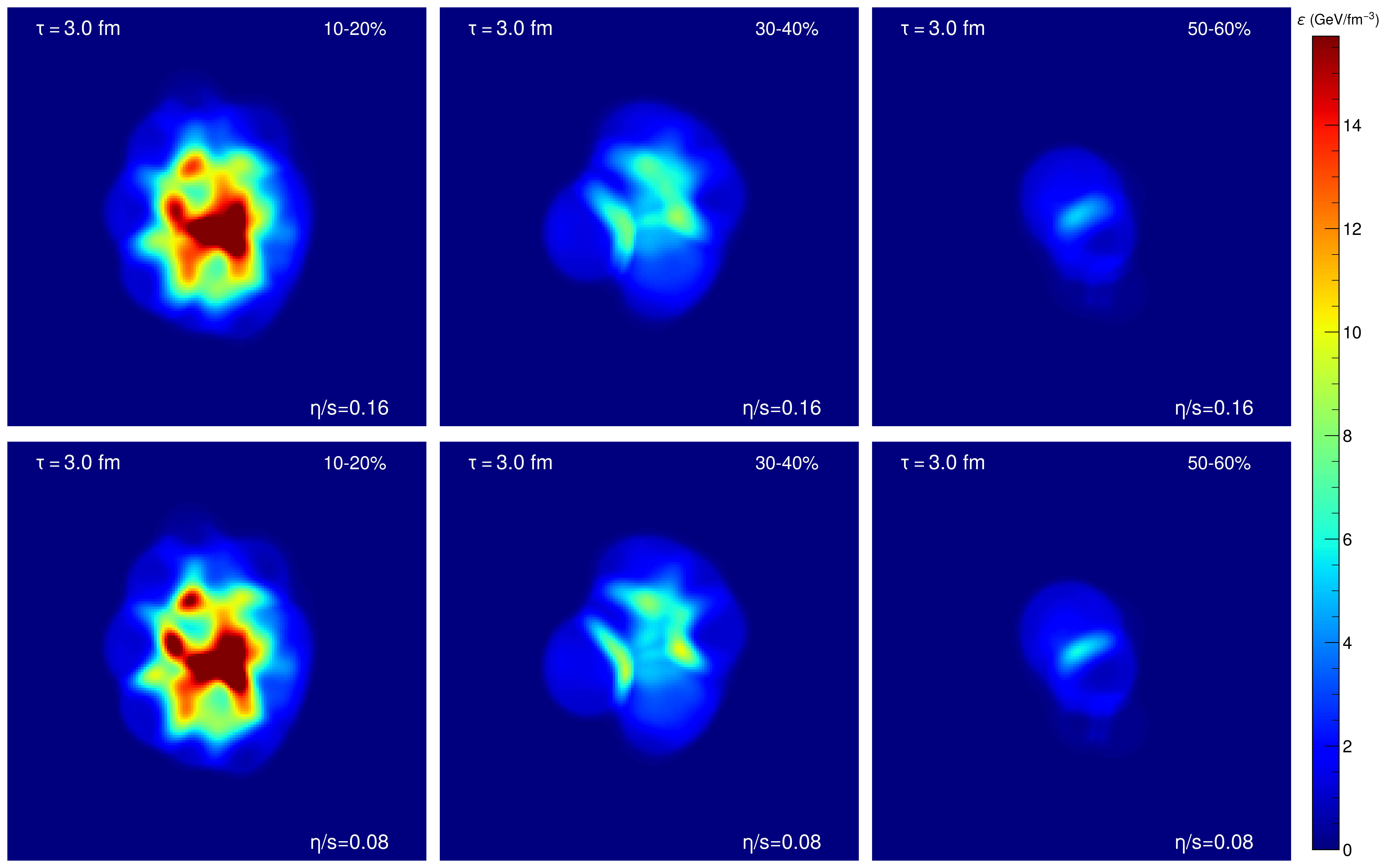}
	\caption{The energy density in the transverse plane at proper time $\tau$=3~fm in Pb+Pb collisions at $\sqrt{s_{NN}}$=5.02 TeV. From left to right are 0-5\%, 10-20\% and 30-40\% centrality classes. The upper panels stands for ${\eta_v/s}$=0.16, while the bottom panel for ${\eta_v/s}$=0.08.}
	\label{tau3p0_evo}
\end{figure*}

\begin{figure*}[t]
	\centering
	\includegraphics[width=\textwidth]{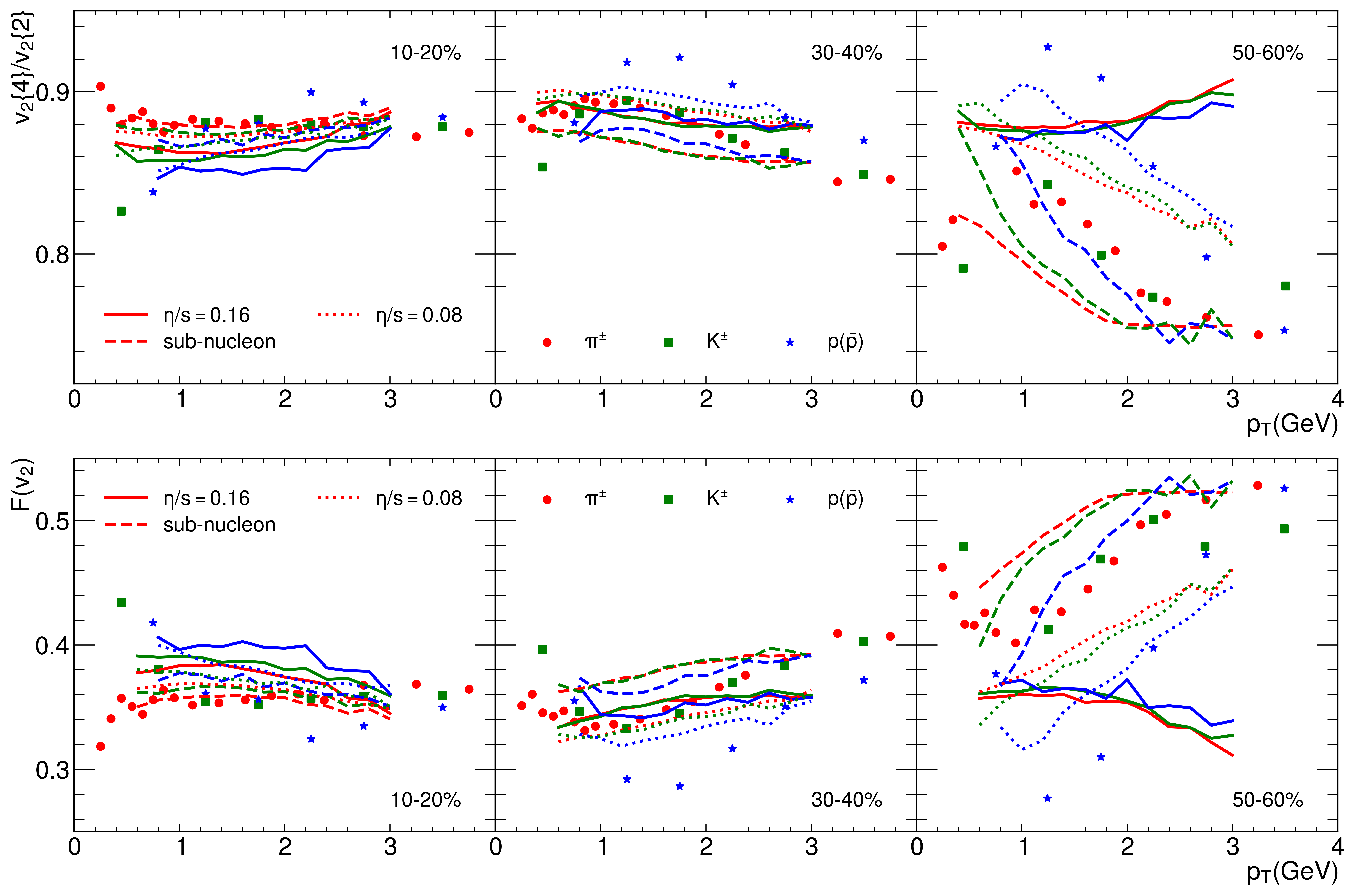}
	\caption{The multi-particle cumulant ratio ${v_2\{4\}/v_2\{2\}}$ and the relative flow fluctuation $F({v_2})$ for identified particles as a function of  transverse momentum $p_T$  in Pb+Pb collisions $\sqrt{s_{NN}}$ = 5.02 TeV for 10-20\%, 30-40\%, and 50-60\% centrality classes. 
	The solid  and dotted line represent the hydrodynamic results with TRENTO initial condition for $\eta_v/s$=0.16 and $\eta_v/s$=0.08, respectively. The blue dashed line represents  the hydrodynamics result with  the sub-nucleon structure in the TRENTO initial condition and $\eta_v/s=0.16$.
	The data are taken from ALICE collaboration~\cite{ALICE:2022zks}.}
	\label{ref_sigma}
\end{figure*}

\begin{figure*}[t]
    \includegraphics[width=\textwidth]{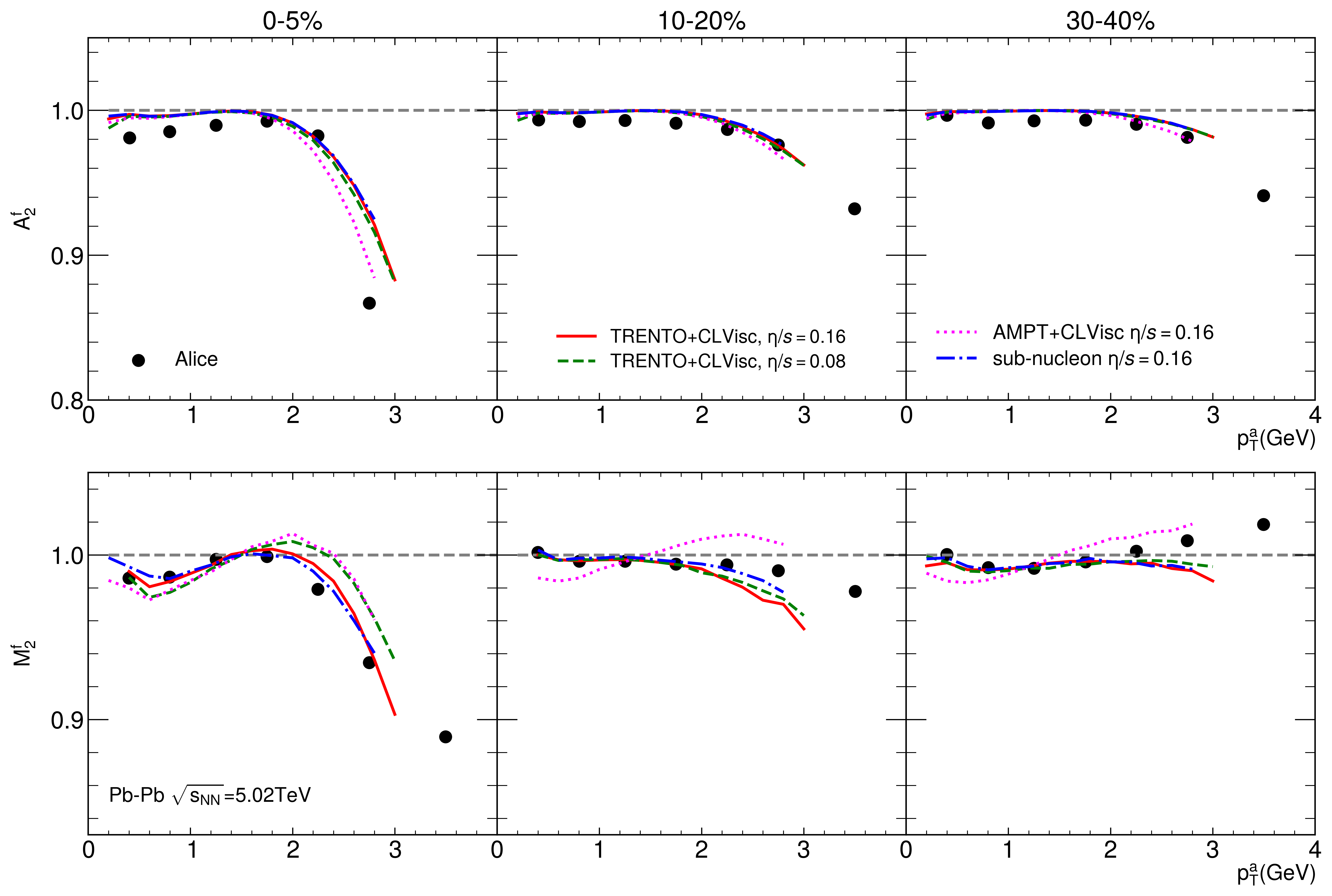}
    \caption{ The flow angle decorrelation $A_2^f$ and flow magnitude decorrelation $M_2^f$  for identified particles as a function of  transverse momentum $p_T$  in Pb+Pb collisions at  $\sqrt{s_{NN}}$ = 5.02 TeV for 0-5\%, 10-20\%, and 30-40\% centrality classes. The solid red line and dashed green line represent the hydrodynamics result with the TRENTO initial condition for $\eta_v/s$=0.16 and $\eta_v/s$=0.08, respectively. The blue dash-dotted line represents the hydrodynamics result with the sub-nucleon structure in the TRENTO initial condition and  $\eta_v/s$=0.16. The pink dotted line stands for the AMPT initial condition and $\eta_v/s$=0.16. The data are taken from ALICE collaboration~\cite{ALICE:2022dtx}.}
    \label{ang_mag}
\end{figure*}

\section{Flow Observables}\label{sec3}

Under the assumptions that the flow vector follows a 2-dimensional Gaussian distribution and the collision geometry gives the dominant effect compared to the flow fluctuations, the two-particle and four-particle correlations can be written as follows~\cite{Voloshin:2008dg},
\begin{equation}  
\begin{aligned}
& v_{{n}}^2\{2\}=\left\langle v_{{n}}\right\rangle^2+\sigma_{{v}_{{n}}}^2, \\
& v_{{n}}^2\{4\} \approx\left\langle v_{{n}}\right\rangle^2-\sigma_{{v}_{{n}}}^2,
\end{aligned}
\end{equation}
where $\left\langle v_{{n}}\right\rangle$ and $\sigma_{{v}_{{n}}}$ are the mean and fluctuation of the anisotropic flow coefficient $v_n$, respectively. 
Using the above formula,  the mean flow $\left\langle v_{{n}}\right\rangle$ and flow fluctuation $\sigma_{{v}_{{n}}}$ can be estimated via $v_{{n}}\{2\}$ and $v_{{n}}\{4\}$ as:
\begin{equation}  
\begin{aligned}
& \left\langle v_{{n}}\right\rangle \approx \sqrt{\frac{v_{{n}}^2\{2\}+v_{{n}}^2\{4\}}{2}}, \\
& \sigma_{{v}_{{n}}} \approx \sqrt{\frac{v_{{n}}^2\{2\}-v_{{n}}^2\{4\}}{2}} .
\end{aligned}
\end{equation}
The relative flow fluctuation $F\left(v_{{n}}\right)$ can be calculated as~\cite{ALICE:2022zks}:
\begin{equation}  
F\left(v_{{n}}\right)=\frac{\sigma_{{v}_{{n}}}}{\left\langle v_{{n}}\right\rangle}.
\end{equation}
Note that the relative flow fluctuation can also be quantified by the ratio $v_2\{4\}/v_2\{2\}$.

Based on the standard multiple cumulant method~\cite{Bilandzic:2010jr}, the $p_T$ differential flow coefficient can be defined via two-particle correlation as follows,
\begin{equation}
\begin{aligned}
	{v_{{n}}\{2\}(p_T)}
	&{=\frac{\langle\langle e^{in(\phi_1^a-\phi_{{2}}}\rangle \rangle}{\sqrt{\langle\langle e^{in(\phi_1-\phi_{{2}})}\rangle\rangle}}
	=\frac{\langle \vec{V}_{{n}}(p_T)\vec{V}_{{n}}^*\rangle}{\sqrt{\langle \vec{V}_{{n}}\vec{V}_{{n}}^*\rangle}}}\\
	&{=\frac{\langle v_{{n}}(p_T)v_{{n}}\cos n(\Psi_{{n}}(p_T)-\Psi_{{n}})\rangle}
	{\sqrt{\langle v_{{n}}^2\rangle}}},
	\end{aligned}
 \end{equation}
where the single bracket denotes the event averages, and the double angle brackets denote the average over both particles and events. $\phi_1^a$ is the azimuthal angles of final particles from some specific $p_T$ range, ${\Psi_{{n}}(p_T)}$ is the corresponding $p_T$ differential event-plane angle, $\phi_1$ and $\phi_2$ are the azimuthal angles of final particles (without $p_T$ constraint), and $\Psi_{{n}}$ stands for $p_T$-integrated event plane angle. 
Due to event-by-event fluctuation, $p_T$ differential event plane ${\Psi_{{n}}(p_T)}$ does not always align with $p_T$ integrated $\Psi_{{n}}$.
Therefore, the above differential flow coefficient includes the decorrelation effect. 
Following the standard procedure, the $\eta$-gap technology is utilized in this work to eliminate the short-range correlation.

One may also define the $p_T$ differential flow coefficient using the four-particle correlation method as follows,
\begin{equation}
v_n\{4\}(p_T)=-\frac{d_n\{4\}(p_T)}{\left(-c_n\{4\}\right)^{3 / 4}},
\end{equation}
where $d_n\{4\}(p_T)$ is related to the fourth-order $p_T$ differential cumulant and $c_n\{4\}$ corresponds to the fourth-order $p_T$ integrated cumulant, as described in Ref.~\cite{Bilandzic:2010jr}.

In order to separate the flow angle and flow magnitude fluctuations in the ${p_T}$-dependent flow vector,  ALICE collaboration~\cite{ALICE:2022dtx} has recently proposed a new set of four-particle correlation functions,
	\begin{equation}\label{ang_form}
	\begin{aligned}
	{A_{{n}}^f}
	&{=\frac{\langle\langle \cos n[\phi_1^a+\phi_{{2}}^a-\phi_3-\phi_4]\rangle\rangle}{\langle\langle \cos n[\phi_1^a+\phi_{{2}}-\phi_3^a-\phi_4]\rangle\rangle}}\\
	&{=\frac{\langle v_{{n}}^2(p_T^a)v_{{n}}^2\cos 2n[\Psi_{{n}}(p_T^a)-\Psi_{{n}}]\rangle}{\langle v_{{n}}^2(p_T^a)v_{{n}}^2\rangle}}\\
	&{\approx \langle \cos 2n[\Psi_{{n}}(p_T^a)-\Psi_{{n}}]\rangle},
	\end{aligned}
	\end{equation}
and
	\begin{equation}\label{mag_form}
	\begin{aligned}
	{M_{{n}}^f}
	&{=\frac{\langle\langle \cos n[\phi_1^a+\phi_{{2}}-\phi_3^a-\phi_4]\rangle\rangle }{ (\langle\langle \cos n[\phi_1^a-\phi_3^a]\rangle\rangle \langle\langle \cos n[\phi_{{2}}-\phi_4]\rangle\rangle)}} \\
	&{\qquad \bigg/\frac{\langle\langle \cos n[\phi_1+\phi_{{2}}-\phi_3-\phi_4]\rangle\rangle }{ \langle\langle \cos n[\phi_1-\phi_{{2}}]\rangle\rangle^2}}\\
	&{=\frac{\langle v_{{n}}^2(p_T^a)v_{{n}}^2\rangle / (\langle v_{{n}}^2(p_T^a)\rangle \langle v_{{n}}^2\rangle)}{\langle v_{{n}}^4\rangle /\langle v_{{n}}^2\rangle^2}},
	\end{aligned}
	\end{equation}
where $A_{n}^f$ ($M_{n}^f$) denotes the decorrelation between $p_T$ differential flow angle (magnitude) and $p_T$ integrated flow. 
If $p_T$ differential flow is the same as $p_T$ integrated flow, then the $A_{n}^f$ and $M_{n}^f$ should be the unity. 
It should be noted that the flow angle decorrelation $A_{n}^f$ should be less than or equal to unity due to the cosine function in the definition. 
However, $M_{n}^f$ does not have such upper limit, due to the fact that it not only involves the covariance between the $p_T$ differential flow and the $p_T$ integrated flow, but also their relative fluctuations. 

\section{Results and Discussion}\label{sec4}

In this section, we present the numerical results for the $p_T$ differential flow coefficients, flow fluctuation, flow angle and magnitude decorrelations in Pb+Pb collisions at $\sqrt{s_{NN}}$=5.02 TeV. These results are based on the (3+1)-dimensional CLVisc hydrodynamics framework with fluctuating  TRENTO and AMPT initial conditions~\cite{Lin:2004en}. We first focus on the  $p_T$ differential flow and flow fluctuations of identified particles in various centrality intervals and study the effects of shear viscosity. Later, we discuss the granularity of initial state fluctuations by considering the nucleon and sub-nucleon structures in the initial condition, respectively. Finally, we demonstrate the hydrodynamic description of flow angle and flow magnitude fluctuations, and their dependence on initial condition model, shear viscosity and sub-nucleon structure.

\subsection{Flow coefficient and flow fluctuation}\label{sec4a}

In Fig.~\ref{v24}, we show ${p_T}$ differential ${v_2\{2\}}$ with rapidity gap $\vert \Delta {\eta} \vert > 0.8$ (upper) and $p_T$ differential ${v_2\{4\}}$ (lower panel) for identified particles (${\pi^\pm, K^\pm, p+\bar{p}}$) in Pb+Pb collisions at $\sqrt{s_{NN}}$=5.02 TeV. 
The subfigures from left to right refer to three centrality classes: 10-20\%, 30-40\%, and 50-60\%, respectively. 
The experimental data from the ALICE collaboration~\cite{ALICE:2022zks} are shown for comparison.
For CLVisc hydrodynamics simulations, the effects of shear viscosity and sub-nucleon structure are shown in the figure. 
First, we observe that hydrodynamics calculation with shear viscosity $\eta_v/s$ = 0.16 can describe the ${v_2\{2\}}$ and ${v_2\{4\}}$ of $\pi^{\pm}$ and $K^{\pm}$ up to 2.5 GeV. 
However, our model overestimates the experiment data of protons at low transverse momentum. This deviation increases from peripheral to central collisions. This can be attributed to the absence of a hadronic afterburner in our calculations, which would produce more blue shift effects for proton, but have mild effects on $\pi^{\pm}$ and $K^{\pm}$. 
Second, we observe that ${v_2\{2\}}$ and ${v_2\{4\}}$ of identified particles shows a weak dependence on the sub-nucleon structure, except ${v_2\{4\}}$ in peripheral collisions. 
In peripheral collisions, the collision systems are smaller and tend to be more influenced by the sub-nucleon structure. 
To further illustrate this, we present in Fig.~\ref{ini-sub} the energy density distribution in the transverse plane at proper time $\tau_0$=0.6 fm in Pb+Pb collisions at $\sqrt{s_{NN}}$=5.02 TeV with and without the sub-nucleon structure. 
One can see that large hot spots split into many small hot spots after considering the sub-nucleon structure. 
Due to the competition among many small hot spots, the underlying geometry of the collision zone does not fully convert into final anisotropic flow after hydrodynamic evolution.

In Fig.~\ref{sigma}, we present the mean value ${\langle v_2\rangle}$ and the standard deviation ${\sigma_{v_2}}$ of the elliptic flow ${v_2}$ as a function of ${p_T}$ for the same particle species, centrality intervals, and model setup as in Fig.~\ref{v24}. 
One can see that the mean flow value ${\langle v_2\rangle}$ and the flow fluctuation ${\sigma_{v_2}}$ exhibit similar ${p_T}$ dependence and mass ordering behavior as the two-particle and four-particle elliptic flow ($v_2\{2\}$ and $v_2\{4\}$). 
The solid curves with shear viscosity $\eta_v/s$=0.16 can better describe the trends of ${\langle v_2\rangle}$ and ${\sigma_{v_2}}$ data up to 2-3 GeV for $\pi^{\pm}$, $K^{\pm}$. 
Similar to $v_2\{2\}$ and $v_2\{4\}$, both ${\langle v_2\rangle}$ and ${\sigma_{v_2}}$ are also suppressed by the shear viscosity. 
This can be understood from Fig.~\ref{tau3p0_evo} which shows the energy density distribution on the transverse plane at proper time $\tau$ = 3 fm in Pb-Pb collisions at $\sqrt{s_{NN}}$=5.02 TeV with different values of shear viscosity: $\eta_v/s=0.16$ and $\eta_v/s=0.08$. 
One can clearly see that the larger shear viscosity has stronger smearing power and therefore results in more isotropic structure. 
In addition, we observe from Fig.~\ref{sigma} that the mean flow ${\langle v_2\rangle}$ is insensitive to the effect of sub-nucleon structure. 
However, the flow fluctuation ${\sigma(v_2)}$ shows very interesting dependence on the sub-nucleon structure. 
In central collisions, ${\sigma(v_2)}$ is suppressed by the sub-nucleon structure, but in peripheral collisions it is actually enhanced by the sub-nucleon structure. 
This can be understood as follows. 
From Fig. \ref{ini-sub}, one can see that for larger systems (i.e., more central collisions), the inclusion of sub-nucleon structure can generate many small hot spots, which can partially cancel the flow developed by each hot spot. 
In contrast, for smaller systems (i.e., more peripheral collisions), the flow pattern can be easily influenced by the sub-nucleon structure (fluctuation).

To further understand the fluctuation of elliptic flow $v_2$, Fig.~\ref{ref_sigma} shows the effect of shear viscosity and sub-nucleon structure on the $p_T$ dependence of the multi-particle cumulant ratio ${v_2\{4\}/v_2\{2\}}$ and the relative fluctuation function $F({v_2})$ in Pb+Pb collisions $\sqrt{s_{NN}}$=5.02 TeV for different centrality classes. 
One can see that in central collisions, ${v_2\{4\}/v_2\{2\}}$ and $F(v_2)$ do not show strong $p_T$ dependence, while in peripheral collisions, the relative flow fluctuations show quite different $p_T$ dependence for different model setups.  
The dashed curves with shear viscosity $\eta_v/s$=0.16 and the sub-nucleon structure in the initial condition model can better describe the trends of ${v_2\{4\}/v_2\{2\}}$ and $F(v_2)$ data up to 2-3 GeV.
If one removes the sub-nucleon structure (as shown by the solid curves), the relative fluctuation becomes smaller, resulting in larger values for ${v_2\{4\}/v_2\{2\}}$ and smaller values for $F(v_2)$. 
In the mean time, the $p_T$ dependence of the relative flow fluctuations also becomes reverse in peripheral collisions.
For the viscosity effect, if one increases the value of shear viscosity to entropy density ratio ($\eta_v/s$), both flow and flow fluctuation become smaller because of the smearing characteristics of the viscosity. 
This leads to different viscosity dependence for relative flow fluctuation in different centrality.
In central collisions, shear viscosity increases the relative flow fluctuation, while in peripheral collisions, the shear viscosity tends to suppress the relative flow fluctuation, especially at large $p_T$ region (the low $p_T$ region is more similar to central collisions).

\subsection{Flow angle and flow magnitude decorrelation}\label{sec4b}

To better understand the flow fluctuation and decorrelation, the ALICE collaboration has proposed two new sets of $p_T$ dependent decorrelation observables (${A_2^f}$  and ${M_2^f}$) based on the four-particle correlation method, which can separate the contributions from flow angle and flow magnitude decorrelation. 
Figture~\ref{ang_mag} presents the flow angle and flow magnitude decorrelation functions ${A_2^f}$  and ${M_2^f}$ as a function of the associated particle transverse momentum  ${p_T^a}$ in Pb+Pb at $\sqrt{s_{NN}} = 5.02$ TeV for several centrality classes (0-5\%, 10-20\%, and 30-40\%). 
The effect of shear viscosity and sub-nucleon structure are shown in the figure. 
We first observe that the deviation of flow angle decorrelation ${A_2^f}$ from unity increases with the transverse momentum $p_T^a$, consistent with experimental data. 
This indicates that the flow angle for higher $p_T$ particles is more deccorrelated from the $p_T$ integrated flow angle. 
This can be understood from the fact that low ${p_T}$ particles mainly originate from the inner part of QGP medium and participate longer hydrodynamic expansion, therefore carry more information about $p_T$ integrated event plane. 
In contrast, high ${p_T}$ particles mainly come from the edges of QGP fireball and experience shorter hydrodynamics evolution, thus carry less information about the overall flow. 
The flow magnitude decorrelation ${M_2^f}$ shows similar ${p_T^a}$ dependence to $A_2^f$ in 0-5\% and 10-20\% centrality classes. 
However, in 30-40\% centrality class, the value of ${M_2^f}$ may beome larger than unity at high $p_T^a$. 
As has been mentioned before, $M_{n}^f$ not only depends on the covariance between the $p_T$ differential flow magnitude and the $p_T$ integrated flow, but also involves their relative fluctuations. Therefore, it is not necessarily bounded to unity. 
$M_{n}^f >1 $ might indicate that a strong positive correlation exists between $p_T$ differential flow magnitude and $p_T$ integrated flow magnitude, and/or the relative fluctuation for $p_T$ differential flow magnitude is very large (at certain $p_T$). 
Another interesting feather is that the decorrelation effect (i.e., the deviation from unity) for both flow angle and flow magnitude is the largest in central collisions, due to no underlying anisotropy for the collision zone. 
For the viscosity effect, we find the flow angle decorrelation $A_2^f$ is not sensitive to shear viscosity. 
The flow magnitude decorrelation ${M_2^f}$ shows stronger dependence on the shear viscosity, especially in most central collisions. 
For the initial condition effect, the flow angle decorrelation $A_2^f$ does not show much dependence on initial condition.
However, the AMPT initial conditions give a completely different $p_T$ dependence for flow magnitude decorrelation ${M_2^f}$ compared to the TRENTO initial conditions in 10-20\% and 30-40\% centrality classes. 
As for the granularity of the initial state, our result indicates that the sub-nucleon structure has negligible effect on flow angle and magnitude decorrelations due to relatively larger system size compared to the size of sub-nucleon fluctuation.

\section{Summary}\label{sec5}

In this work, we have performed a systematic study on elliptic flow coefficients, flow fluctuation and flow decorrelation in Pb+Pb collisions at $\sqrt{s_{NN}}$=5.02 TeV based on the (3+1)-Dimensional CLVisc hydrodynamics framework with the TRENTO (and AMPT) initial conditions. 
In particular, we have studied the effect of shear viscosity and sub-nucleon structure on these flow observables.  

Our numerical results with shear viscosity $\eta_v/s=0.16$ can reasonably describe the experimental data on two-particle $v_2\{2\}$ and four-particle $v_2\{4\}$, mean flow ${\langle v_2\rangle}$ and flow fluctuation ${\sigma_{v_2}}$ for $\pi^{\pm}$ and $K^{\pm}$ particles. 
For protons, our result overestimates the elliptic flow and flow fluctuation due to the absence of hadronic afterburner in the simulation. 
Shear viscosity tends to suppress the elliptic flow coefficient and flow fluctuation, due to its smearing effect on the local density fluctuation. 
In addition, the flow coefficients appear to be insensitive to the sub-nucleon structure, while the flow fluctuation does depend on the sub-nucleon structure: it tends to be suppressed by the sub-nucleon structure in central collisions but enhanced in peripheral collisions. 
Our result including the sub-nucleon structure can better describe the experimental data on the relative flow fluctuation observables, such as ${v_2\{4\}/v_2\{2\}}$ and $F({v_2})$, measured by the ALICE Collaboration  at the LHC.

We have further studied the effects of the shear viscosity, the sub-nucleon structure and the initial condition on flow angle and flow magnitude decorrelations (${A_2^f}$, ${M_2^f}$) using four-particle correlation method as proposed by the ALICE Collaboration. 
We found that the decorrelation effect for both flow angle and flow magnitudes is largest in most central collisions. Also, the flow angle decorrelation effect increases with the transverse momentum.
In addition, the flow angle decorrelation is insensitive to shear viscosity and only slightly depends on initial condition in most central collisions, whereas the flow magnitude decorrelation shows sensitivity to both shear viscosity and initial condition. 
In particular, the AMPT initial condition gives very different behavior for the flow magnitude decorrelation as compared to the TRENTO initial condition. 

In the future, it would be interesting to include the hadronic afterburner in the simulation, which will allow us to investigate additional sources of flow fluctuation and decorrelation. Additionally, performing a Bayesian analysis based on calibrations including flow fluctuation and decorrelation observables can further help to constrain the transport properties and initial structure of the QGP produced in relativistic heavy-ion collisions.

\section{ACKNOWLEDGMENTS}

This work is supported in part by Natural Science Foundation of China (NSFC) under Grant Nos. 12225503, 11890710, 11890711 and 11935007. Some of the calculations were performed in the Nuclear Science Computing Center at Central China Normal University (NSC$^3$), Wuhan, Hubei, China.

\bibliographystyle{h-physrev5} 
\bibliography{refs}   
\end{document}